\documentclass[aip, jcp, amsmath,amssymb, reprint, footinbib]{revtex4-1}

\usepackage{graphicx}
\usepackage{dcolumn}
\usepackage{bm}

\begin{document}

\title{Generalized Quantum Master Equations In and Out of Equilibrium: When Can One Win?}

\author{Aaron Kelly}
\thanks{These authors contributed equally to this work.}
\affiliation{Department of Chemistry, Stanford University, Stanford, California, 94305, USA}
\author{Andr\'{e}s Montoya-Castillo}
\thanks{These authors contributed equally to this work.}
\affiliation{Department of Chemistry, Columbia University, New York, New York, 10027, USA}
\author{Lu Wang}
\affiliation{Department of Chemistry and Chemical Biology, Rutgers University, Piscataway, New Jersey, 08854, USA}
\author{Thomas E. Markland}
\email{tmarkland@stanford.edu}
\affiliation{Department of Chemistry, Stanford University, Stanford, California, 94305, USA}

\date{\today}

\begin{abstract}
Generalized quantum master equations (GQMEs) are an important tool in modeling chemical and physical processes. For a large number of problems it has been shown that exact and approximate quantum dynamics methods can be made dramatically more efficient, and in the latter case more accurate, by proceeding via the GQME formalism. However, there are many situations where utilizing the GQME approach seems to offer no advantage over a direct evaluation of the property of interest. Here we provide a more detailed understanding of the conditions under which these methods will offer benefits. In particular, we derive exact expressions for the memory kernel for systems both in and out of equilibrium, and show the conditions under which these expressions will be guaranteed to return a result identical to that obtained from direct simulation. We also show the conditions which approximate methods must satisfy if they are to offer different results when used in conjunction with the GQME formalism. These exact analytical results thus provide new insights as to when proceeding via the GQME approach can improve the accuracy or efficiency of simulations.
\end{abstract}

\maketitle

Generalized quantum master equations (GQMEs) provide a formal framework to describe the time evolution of observables and correlation functions in complex, many-body, systems based on the projection operator method.\cite{Nakajima1958,Zwanzig1960a,Mori1965} The generalized master equation formalism has found extensive use both in allowing efficient and accurate calculations of material properties, including diffusion constants of liquids,\cite{Martin1968, Harp1970, Levesque1970, Boon1967,Nee1970} density fluctuations in glasses,\cite{Rabani2004,Rabani2005,Markland2011,Markland2012} and structural relaxation in polymers.\cite{Felderhof1975,Schweizer1989} In addition, it has also been heavily exploited as an analysis tool to uncover the inherent timescales in complex chemical systems,\cite{Chandler1978} as a dimensionality reduction technique in the development of coarse-grained molecular models,\cite{Hijon2010} and more broadly in areas such as meteorological and financial time-series analysis and optimal prediction methods.\cite{Chorin2000,Schmitt2006,Niemann2008,Venturi2014} The central quantity in the GQME formalism is the the memory kernel, which encodes the effect of the projected dynamical degrees of freedom on the observable. However, the standard expressions for the memory kernel contain projected dynamical quantities that are impractical to simulate directly. This has led to the development of a number of ways to approximate the memory kernel that allow it to be recast in terms of unprojected dynamical quantities,\cite{Berne1966,Berne1990} assumed functional forms,\cite{Chen2010a} or a given (perturbative or Markovian) limit.\cite{Bloch1957,Redfield1965,Dekker1987,Leggett1987,Weiss,Cheng2005}

Just over a decade ago, Shi and Geva derived a formally exact representation for the memory kernel of the Nakajima-Zwanzig GQME that requires only projection-free input.\cite{Shi2003} This representation opened the door to using either numerically exact or approximate methods to simulate the memory kernel. For exact treatments, whose computational cost increases severely with propagation time, exploiting the rapid decay of the memory kernel has been shown to allow for significant gains in the efficiency of simulating charge and energy transport in the condensed phase.\cite{Shi2003,Shi2004a,Shi2004,Zhang2006a,Cohen2011,Cohen2013, Cohen2013a,Wilner2013,Wilner2014,Wilner2015,Kidon2015a}  When approximate dynamics are used, such as those arising from the quantum-classical and semiclassical hierarchies,\cite{McLachlan1964,Tully1971,Miller2001a,Miller2006,Kapral2015} significant improvements in both accuracy and efficiency have been achieved when used to calculate the memory kernel compared to their direct application.\cite{Shi2004a,Shi2004,Kelly2013, Kelly2015,Pfalzgraff2015,Montoya2016a}  However, it has been shown that such improvements sensitively depend on how one calculates the projection-free partial kernels that are used to construct the memory kernel.\cite{Montoya2016a} These observations naturally raise questions as to why this is and when proceeding via the projection operator formalism will be advantageous. 

Here we show, for both equilibrium and nonequilibrium systems, the conditions under which proceeding via the GQME formalism yields results that are guaranteed to be identical to the original dynamics used in the projection free input, and suggest ways in which this limitation can be overcome. To achieve this we show how the memory kernel governing the evolution of equilibrium and nonequilibrium systems can be exactly cast in terms of unprojected correlation functions, which can be straightforwardly simulated using either exact or approximate methods. By analyzing these expressions we derive the necessary requirements for an approximate dynamics to yield the same results when used directly and as an approximation to the memory kernel in the GQME approach. These results thus provide insights into when GQME methods might allow for improvement in accuracy or efficiency in equilibrium and nonequilibrium situations in a diverse set of systems.

To begin we consider the Nakajima-Zwanzig GQME,\cite{Nakajima1958,Zwanzig1960a} which provides a route for describing the reduced dynamics of systems out of equilibrium. In this scheme the total system is decomposed into two parts: the system, which consists of all the degrees of freedom of interest in the problem, and the bath, which comprises the remaining degrees of freedom. For equilibrium systems the Mori approach is typically the preferred formulation.\cite{Mori1965} However, the Mori formalism, which renders distinction between system and bath unnecessary, is general and can also be used for equilibrium and nonequilibrium problems. Indeed, as observed in Ref.~\onlinecite{Montoya2016a}, both the Mori and Nakajima-Zwanzig approaches can be written in a unified formalism stemming from the projected equation of motion for the propagator,
	\begin{equation}\label{Eq:Prop}
	\frac{d}{dt}e^{i\mathcal{L}t} = e^{i\mathcal{L}t}i\mathcal{L} = e^{i\mathcal{L}t}(\mathcal{P}+\mathcal{Q})i\mathcal{L}
	\end{equation}
where the Liouville operator is $\mathcal{L} = \frac{1}{\hbar}[ \hat{H}, \cdot]$, $\hat{H}$ is the Hamiltonian operator, $\mathcal{P}$ is a projection operator and $\mathcal{Q} = \mathbf{1} - \mathcal{P}$ is the complementary projection operator. 

Central to the Mori approach is the appropriate choice of projection operator, $\mathcal{P}$.  In the following, we assume that it takes the form, 
    \begin{equation}
        \mathcal{P} = |\mathbf{A} )( \mathbf{A}|, 
    \end{equation}
where $\mathbf{A}$ contains a subset of observables which are of particular interest.  The definition of the inner product is chosen such that $(\mathbf{A}|\mathbf{A}) = \mathbf{1}$, thus satisfying the idempotency condition, $\mathcal{P}^2 = \mathcal{P}$. Consequently, the complementary projector, $\mathcal{Q}$, is by construction orthogonal to the subspace defined by $\mathcal{P}$. 

Using the Dyson operator identity 
    \begin{equation}\label{Eq:DysonDecomp}
       e^{i\mathcal{L}t} =  e^{i\mathcal{Q}\mathcal{L}t} + \int_0^t d\tau\  e^{i\mathcal{L}(t-\tau)}(\mathcal{P}i\mathcal{L})e^{i\mathcal{Q}\mathcal{L}\tau},
    \end{equation}
to expand the second term in the second equality of Eq.~(\ref{Eq:Prop}) yields
    \begin{equation}\label{Eq:PropagatorEOM}
    \begin{split}
        \frac{d}{dt}e^{i\mathcal{L}t} &= ie^{i\mathcal{L}t}\mathcal{P}\mathcal{L} + i \mathcal{Q}e^{i\mathcal{L}\mathcal{Q}t}\mathcal{L} \\
	    &\qquad -  \int_0^{t}d\tau\ e^{i\mathcal{L}(t-\tau)}\mathcal{P}\mathcal{L}\mathcal{Q}e^{i\mathcal{L}\mathcal{Q}\tau}\mathcal{L}.
    \end{split}
    \end{equation}

Restricting our attention to correlation functions defined by the inner product of the elements constituting $\mathcal{P}$, the equation of motion for the propagator in Eq.~(\ref{Eq:PropagatorEOM}) yields the following Mori-type GQME for the correlation function $\mathcal{C}(t) = (\mathbf{A}|\mathbf{A}(t))$, 
    \begin{equation} \label{Eq:EOM_C}
        \dot{\mathcal{C}}(t) = \mathcal{C}(t)\dot{\mathcal{C}}(0) - \int_0^t d\tau\ \mathcal{C}(t-\tau)\mathcal{K}(\tau), 
    \end{equation}
and the memory kernel, $\mathcal{K}(t)$, takes the form, 
    \begin{equation}\label{Eq:K}
        \mathcal{K}(t) = (\mathbf{A}|\mathcal{L}\mathcal{Q} e^{i\mathcal{Q}\mathcal{L}t} \mathcal{Q}\mathcal{L} |\mathbf{A}).
    \end{equation}
Direct evaluation of the memory kernel in Eq.~(\ref{Eq:K}) is problematic, as it requires the action of the projected propagator, $e^{i\mathcal{Q}\mathcal{L}t}$.  

To circumvent the difficulty of the projected propagator, the Dyson identity, Eq.~(\ref{Eq:DysonDecomp}), can be used to obtain a self-consistent expansion of the memory kernel, 
    \begin{equation}\label{Eq:KSC}
        \mathcal{K}(t) = \mathcal{K}_{1}(t) + \int_0^t d\tau\ \mathcal{K}_{3}(t-\tau)\mathcal{K}(\tau),
    \end{equation}
where the partial (auxiliary) kernels, 
    \begin{align}
        \mathcal{K}_{1}(t) &= (\mathbf{A}| \mathcal{L}\mathcal{Q}e^{i\mathcal{L}t}\mathcal{Q}\mathcal{L}|\mathbf{A}) , \label{Eq:K1}\\
	\mathcal{K}_{3}(t) &= -i(\mathbf{A}| \mathcal{L}\mathcal{Q}e^{i\mathcal{L}t} |\mathbf{A}) \label{Eq:K3},
    \end{align}
no longer require the use of projected dynamics. The labels for the partial kernels of $1$ and $3$ are chosen so as to be consistent with earlier work.\cite{Shi2003,Shi2004a,Kelly2013, Kelly2015,Pfalzgraff2015,Montoya2016a} In principle, the memory kernel $\mathcal{K}(t)$ can be obtained by generating $\mathcal{K}_{1}(t)$ and $\mathcal{K}_{3}(t)$ from simulation and solving Eq.~(\ref{Eq:KSC}) numerically. Depending on the choice of projection operator and definition of the inner product, one can specialize this result to equilibrium correlation functions or equilibrium population dynamics.

In the equilibrium case the Kubo-transformed correlation function is obtained by defining the inner product as
    \begin{equation}\label{Eq:EquilInnerProduct}
        (\mathbf{A}|\mathcal{O}|\mathbf{A}) \equiv \int_0^{\beta}d\lambda\  \mathrm{Tr}[\rho_{eq} \mathbf{A}^{\dagger}(0) \mathcal{O}\mathbf{A}(i\lambda)] \cdot \chi_{AA}^{-1},
    \end{equation}
where $\mathcal{O}$ is a general superoperator in Liouville space, $\rho_{eq} = Z^{-1} e^{-\beta \hat{H}}$ is the canonical density operator, $Z = \mathrm{Tr}[e^{-\beta \hat{H}}]$ is the partition function, $\beta = 1/k_BT$ is the inverse of the thermal energy, and $\chi_{AA} = \mathrm{Tr}[\rho \mathbf{A}^{\dagger}(0) \mathbf{A}(0)]$.  Commonly, the elements of the vector $\mathbf{A}$ are chosen to consist of a dynamical variable, $\mathbf{a}$, which is a function of the coordinates and momenta of the system, and its time derivative, $\dot{\mathbf{a}} = i\mathcal{L}\mathbf{a}$. For example, in the case of diffusion $\mathbf{a}$ could be chosen to be the position or velocity of some or all of the particles, and for infra-red spectroscopy as the system dipole moment. 
	
In nonequilibrium cases the inner product may be defined as\cite{Montoya2016a} 
\begin{equation}
    (\mathbf{A}|\mathcal{O}|\mathbf{A})_{nm} \equiv  \mathrm{Tr}[\mathcal{R}_B A_n^{\dagger} \mathcal{O}A_m],
\end{equation}
where the set $\{A_n\}$ spans a limited subspace of total Hilbert space of the system, and $\mathcal{R}_B$ is an operator that belongs to the complementary space, which is conventionally denoted as the bath.  The normalization condition on $\mathcal{R}_B$ requires that the trace over the bath degrees of freedom yield unity, $\mathrm{Tr}_B[\mathcal{R}_B] = 1$.  The elements of $\mathbf{A}$ can, for example, be chosen such that $\{A_n\}$ spans all outer products of the system states and $\mathcal{R}_B = e^{\beta H_B}/\mathrm{Tr}_B[e^{\beta H_B}]$ corresponds to the canonical distribution for the bath degrees of freedom. Such a choice for the projector provides access to nonequilibrium population and coherence dynamics of the system, and has been used in the context of the spin-boson model,\cite{Shi2003,Shi2004a,Kelly2013, Kelly2015,Montoya2016a,Wilner2015} and in a wide class of quantum impurity models to access site occupation dynamics in the presence of one or more noninteracting fermionic or bosonic baths.\cite{Cohen2011,Cohen2013,Cohen2013a,Wilner2013,Wilner2014,Kidon2015a}

The expressions in Eqs. (\ref{Eq:KSC}), (\ref{Eq:K1}), and (\ref{Eq:K3}) have been shown to improve the accuracy of the dynamics produced by a number of approximate methods\cite{Shi2004a,Kelly2013,Pfalzgraff2015,Kelly2015,Montoya2016a} when used in the GQME formalism. However, one can take these exact expressions and manipulate them using relationships that are exactly satisfied in quantum mechanics to obtain alternative expressions that, when approximated, can yield different results. Here we show the manipulations that one can make that can be proved to rigorously return a result that is identical to that obtained using the approximate method directly, irrespective of the method employed. Extension of this analysis reveals the conditions that must be satisfied by an approximate method to guarantee that the same result is obtained no matter which route, or way of formally rewriting the partial kernels, is taken.

To derive exact expressions for the partial kernels which, when used to obtain the dynamics using Eqs.~(\ref{Eq:EOM_C}) and (\ref{Eq:KSC}), are guaranteed give the same result as a direct simulation, regardless of the dynamics approach employed, one begins by expanding the complementary projection operator $\mathcal{Q}$ and applying the Liouville operators.\cite{Montoya2016a} Carrying out the former operation for $\mathcal{K}_{3}$ (Eq.~(\ref{Eq:K3})) yields,
\begin{equation}
   \mathcal{K}_{3}(t) = -i(\mathbf{A}| \mathcal{L}e^{i\mathcal{L}t} |\mathbf{A})+i(\mathbf{A}| \mathcal{L}|\mathbf{A} )( \mathbf{A}|e^{i\mathcal{L}t} |\mathbf{A}). 
\end{equation}
The Liouville operator, $\mathcal{L}$, can then be applied backwards on the static part or forwards to generate the time derivative. Doing the latter allows the partial kernel to be written purely as a function of $\mathcal{C}(t)$ and its time derivatives, 
\begin{equation}
    \mathcal{K}_{3}(t) = -\dot{\mathcal{C}}(t)+ \dot{\mathcal{C}}(0)\mathcal{C}(t).\label{eq:K3_new}
\end{equation}
Proceeding similarly for $\mathcal{K}_{1}$ gives,
\begin{equation}
	\begin{split} \label{eq:K1_new}
	\mathcal{K}_{1}(t) &= (\mathbf{A}| \mathcal{L}e^{i\mathcal{L}t}\mathcal{L}|\mathbf{A}) - \dot{\mathcal{C}}(0)\mathcal{C}(t)\dot{\mathcal{C}}(0)\\
	&\qquad +i \dot{\mathcal{C}}(0)(\mathbf{A}| e^{i\mathcal{L}t}\mathcal{L}|\mathbf{A}) +i (\mathbf{A}| \mathcal{L}e^{i\mathcal{L}t}|\mathbf{A})\dot{\mathcal{C}}(0) \\	
	&= (\mathbf{A}| e^{i\mathcal{L}t}\mathcal{L}^2|\mathbf{A}) - \dot{\mathcal{C}}(0)\mathcal{C}(t)\dot{\mathcal{C}}(0)\\
	&\qquad +i \{\dot{\mathcal{C}}(0), (\mathbf{A}| e^{i\mathcal{L}t}\mathcal{L}|\mathbf{A})\} \\	
	&= -\ddot{\mathcal{C}}(t) + \{\dot{\mathcal{C}}(0), \dot{\mathcal{C}}(t)\} - \dot{\mathcal{C}}(0)\mathcal{C}(t)\dot{\mathcal{C}}(0),
	\end{split}
\end{equation}
where in the second and last equalities the braces denote the anticommutator.

Due to the convolution-based structure of the equations relating the full memory kernel to the partial memory kernels, it is particularly convenient to consider its Fourier-Laplace representation. The Fourier-Laplace transform of $\mathcal{C}(t)$ is defined as,
    \begin{equation}
	\mathcal{C}(\omega) = \int_0^{\infty} dt\ e^{i\omega t} \mathcal{C}(t),
    \end{equation}
and its n$^{th}$ time-derivative, $\mathcal{C}^{(n)}(t) \equiv \frac{d^{n}}{dt^{n}}\mathcal{C}(t)$, is
	\begin{equation}
    \mathcal{C}^{(n)}(\omega) = (-i\omega)^n \mathcal{C}(\omega) - \sum_{k=1}^n (-i\omega)^{n-k} \mathcal{C}^{(k-1)}(t=0).
    \end{equation}
Application of the Fourier-Laplace transform to Eq.~(\ref{Eq:KSC}) gives,
    \begin{equation}\label{Eq:K_omega}
    \mathcal{K}(\omega) = [\mathbf{1} - \mathcal{K}_{3}(\omega)]^{-1}\mathcal{K}_{1}(\omega)
    \end{equation}
and to Eqs.~(\ref{eq:K1_new}) and (\ref{eq:K3_new}) yields, 
    \begin{align}
    \mathcal{K}_{1}(\omega) &= - \Omega(\omega)[\mathbf{1} +  \mathcal{C}(\omega)\Omega(\omega)],\label{Eq:K1_fourier} \\
    \mathcal{K}_{3}(\omega) &= \mathbf{1} + \Omega(\omega)\mathcal{C}(\omega),\label{Eq:K3_fourier}
    \end{align}
where $\Omega(\omega) = i\omega + \dot{\mathcal{C}}(t=0)$.
Fourier-Laplace transforming and rearranging the equation of motion for the correlation function in Eq. (\ref{Eq:EOM_C}) results in
    \begin{equation} \label{Eq:EOM_freq}
    \mathcal{K}(\omega)  = \mathcal{C}(\omega)^{-1}[\mathbf{1} + \mathcal{C}(\omega)\Omega(\omega)],
    \end{equation}
where we have used $\mathcal{C}(t=0) = \mathbf{1}$, which follows from the idempotency property of the projector. 

Using the Fourier-Laplace transforms of the memory kernels in Eqs.~(\ref{Eq:K1_fourier}) and (\ref{Eq:K3_fourier}) in Eq.~(\ref{Eq:K_omega}) and inserting the result into the left hand side of Eq.~(\ref{Eq:EOM_freq}) returns the identity $\mathcal{C}(\omega)=\mathcal{C}(\omega)$. Hence, when one uses expressions for the partial kernels that depend only on the original correlation function, $\mathcal{C}(t)$, and its time-derivatives, one is certain to recover exactly the same result as would be obtained from a direct application of the dynamics used to calculate the partial kernels. This result therefore proves that no accuracy benefit can be obtained through the GQME for any approximate method using Eqs.~(\ref{eq:K3_new}) and (\ref{eq:K1_new}).\cite{Note1}More explicitly, using manipulations that are exactly satisfied by quantum mechanics to recast the partial kernels in terms of the original correlation function and its time derivatives removes any potential benefit for gains in accuracy by proceeding via the GQME formalism. 

It should be noted that all the expressions given for the partial memory kernels in the preceding equations (i.e., Eqs. (\ref{Eq:K1}) and (\ref{Eq:K3}) as well as Eqs. (\ref{eq:K1_new}) and (\ref{eq:K3_new})) are guaranteed to give the same result as a direct application when exact quantum dynamics is used to generate them. However, if the partial kernels are shorter-lived than the correlation function, one can still obtain significant efficiency gains when using numerically exact methods, which scale poorly with simulation time.

The fact that Eqs.~(\ref{eq:K3_new}) and (\ref{eq:K1_new}) cannot be used to obtain an increase in accuracy for any approximate method begs the question: What are the conditions that an approximate method must satisfy to guarantee that the same result will be obtained by using Eqs.~(\ref{Eq:K1}) and (\ref{Eq:K3}) as  Eqs.~(\ref{eq:K1_new}) and (\ref{eq:K3_new})? Identifying such conditions allows the identification of the classes of approximate methods which \textit{cannot} gain accuracy benefits by combination with the GQME formalism. Indeed, we can immediately see that approximate methods only need to satisfy two such conditions. The first is that  
\begin{equation}\label{Eq:Condition1}
	\frac{d}{dt} \mathcal{C} (t) = ( \mathbf{A} | e^{i\mathcal{L'}t}[i\mathcal{L'} \mathbf{A}] ),
\end{equation}
where $\mathcal{L}'$ indicates a Liouville operators corresponding to an approximate dynamics. Explicitly, this condition requires that the correlation function of $\mathbf{A}$ and the operator resulting from the action of the Liouvillian acting on $\mathbf{A}$ is equivalent to the time-derivative of the original correlation function.  While the equality in Eq.~(\ref{Eq:Condition1}) is trivially maintained for numerically exact methods, the same is not necessarily true for approximate methods. The second condition requires that the approximate Liouville operator commutes with the propagator, 
\begin{equation}\label{Eq:Condition2}
	[\mathcal{L'}, e^{i\mathcal{L'}t}] = 0. 
\end{equation}
These conditions are all that are required of an approximate method to perform the manipulations in Eq.~(\ref{eq:K1_new}) and likewise to obtain Eq.~(\ref{eq:K3_new}) from Eq.~(\ref{Eq:K3}).

In the case of equilibrium correlation functions, where the Liouville operator commutes with the canonical density matrix, Eq.~(\ref{Eq:Condition2}) is equivalent to time-translational invariance. For example, dynamics methods such as CMD,\cite{Jang1999} RPMD,\cite{Craig2004,rpmd_review} and purely classical mechanics satisfy these properties and thus are guaranteed not to obtain increased accuracy when used to approximate equilibrium correlation functions via the GQME approach. However, for nonequilibrium correlation functions, and for methods that do not satisfy these conditions, one can expect the dynamics resulting from the GQME to differ from the direct application of the approximate method, as has been shown in previous work where significant gains in both efficiency and accuracy have been achieved.\cite{Shi2004a,Shi2004,Kelly2013,Kelly2015,Pfalzgraff2015,Montoya2016a} 
    
In this paper we have provided a set of exact expressions for the memory kernel in the Mori and Nakajima-Zwanzig GQME formalisms. These expressions for the memory kernel can be simulated directly using exact or approximate dynamics methods. We have shown that when the memory kernel is formally cast in terms of the original correlation function and its time derivatives, the GQME is guaranteed to return the same result as that obtained from a direct simulation, regardless of the method used to construct the kernel. We have further specified the criteria that an approximate dynamics method must satisfy to produce GQME dynamics that are identical to the direct dynamics, independent of the way the partial kernels are expressed. In cases where no such benefit in accuracy can be obtained, one may still expect the GQME scheme to yield improved efficiency if the memory kernels are short-lived. Conversely, violation of the criteria in Eqs.~(\ref{Eq:Condition1}) and (\ref{Eq:Condition2}) imply that the resulting GQME dynamics will in general be different from direct simulation of the correlation functions. Previous studies\cite{Shi2004a,Kelly2013,Kelly2015,Pfalzgraff2015,Montoya2016a} have indeed confirmed this fact, showing that the GQME approach is capable of yielding significant gains both in efficiency and accuracy when one does not invoke the formal recasting of the partial kernels in Eqs.~(\ref{eq:K3_new}) and (\ref{eq:K1_new}). These insights thus outline a path for future applications of the GQME formalism by allowing one to assess the benefits that it may afford. 

\acknowledgments
We greatly thank David Reichman for numerous insightful discussions. We also thank Will Pfalzgraff for helpful comments and a thorough reading of this manuscript, Mariana Rossi for helpful discussions and numerical verification of some of these analytic results, and Hsing-Ta Chen for useful conversations. This material is based upon work supported by the U.S. Department of Energy, Office of Science, Office of Basic Energy Sciences under Award Number DE-SC0014437. T.E.M also acknowledges support from a Cottrell Scholarship from the Research Corporation for Science Advancement and an Alfred P. Sloan Research fellowship. A.K and L.W. acknowledge postdoctoral fellowships from the Stanford Center for Molecular Analysis and Design.

\end{document}